\documentclass[12pt,a4paper]{article}

\usepackage{latexsym,amsmath,ifthen,ramk}

\newcommand{\?}{\hspace{0.1em}}
\newcommand{\Fv}{F_v}
\newcommand{\Fvhut}{\hat{F}_v}
\newcommand{\Dva}{D_v a}
\newcommand{\Dvahut}{D_v \hat{a}}

\newcommand{\diag}{\rm diag}
\newcommand{\haf}{{\textstyle\frac{1}{2}}}
\newcommand{\kwa}{{\textstyle\frac{1}{4}}}
\newcommand{\br}{\langle}
\newcommand{\ke}{\rangle}
\newcommand{\sss}{\scriptscriptstyle}
\newcommand{\bea}{\begin{eqnarray}}
\newcommand{\eea}{\end{eqnarray}}
\newcommand{\fs}{\; .}
\newcommand{\co}{\; ,}
\newcommand{\al}{&\!\!}
\newcommand{\eff}{{e\hspace{-0.1em}f\hspace{-0.18em}f}}
\newcommand{\QCD}{\mbox{\tiny Q\hspace{-0.05em}CD}}
\newcommand{\indV}{{\mbox{\tiny V}}}
\newcommand{\indA}{{\mbox{\tiny A}}}
\newcommand{\indL}{{\mbox{\tiny L}}}
\newcommand{\indR}{{\mbox{\tiny R}}}
\newcommand{\indU}{{\scriptscriptstyle U}}
\newcommand{\WZW}{{\mbox{\tiny{WZW}}}}
\newcommand{\nc}{N_{\!c}}
\newcommand{\nf}{N_{\!f}}
\newcommand{\qbar}{\overline{\rule[0.42em]{0.4em}{0em}}\hspace{-0.5em}q}
\newcommand{\non}{\nonumber}

\begin{document}

\begin{titlepage}
\begin{flushright}BUTP-00/23\\hep-ph/0011377\end{flushright}
\vspace*{1cm}
\begin{center}
{\LARGE\bf Anomalies and WZW-term of\\\vspace{0.3em}two-flavour QCD}

\vspace{0.8cm}
Roland Kaiser\\Institute for Theoretical Physics,
University of Bern,\\  
Sidlerstr. 5, CH--3012 Bern, Switzerland\\
E-mail: kaiser@itp.unibe.ch

\vspace{0.3cm}
November 2000

\vspace{0.6cm}
\begin{abstract}
The U(2)$_\indR\times$U(2)$_\indL$ symmetry of QCD with two massless
flavours is subject to anomalies which affect
correlation functions involving the singlet currents $A_\mu^0 $ or
$V_\mu^0$. These are relevant for $\pi\gamma$-interactions,
because -- for two flavours -- the electromagnetic
current contains a singlet piece. We give the effective 
Lagrangian required  
for the corresponding low energy analysis 
to next-to-leading order, without invoking an expansion in the
mass of the strange quark. In particular, the Wess-Zumino-Witten
term that 
accounts for the two-flavour anomalies within the effective theory is
written down 
in
closed form. 
\end{abstract}

\vspace{2cm}
\footnotesize{\begin{tabular}{ll}
{\bf{Pacs:}}$\!\!\!\!$&11.30.Rd, 12.38.Aw, 12.39.Fe, 14.40.-n, 14.65.Bt\\
{\bf{Keywords:}}$\!\!\!\!$& QCD, chiral symmetry, anomalies, effective
Lagrangian,    
\\& chiral perturbation
theory
\end{tabular}}
\vspace{1.5cm}

\rule{33em}{0.02em}\\
{\footnotesize Work supported in part by Schweizerischer Nationalfonds}  
\end{center}
\end{titlepage}

\clearpage
\setcounter{page}{2}
\section{Introduction}
\label{intro}

It is well known that chiral transformations are, in general,
afflicted with 
anomalies 
\cite{Adler:1969gk,Bardeen,WZ}. In quantum chromodynamics, the most prominent
example of an 
anomalous process is the decay $\pi^0 \to \gamma \gamma $. At low energies,
the properties of the pions may be  
  investigated 
  by means 
  of chiral
perturbation theory \cite{Weinberg Physica,GL SU(2)}. 
In the effective theory, the anomalies are accounted for by the
Wess-Zumino-Witten term \cite{WZ,GL SU(3),W,WZW,KT}. In connection with the
decay 
$\pi^0 \to \gamma \gamma $, it is therefore remarkable to note that the
effective theory as constructed in ref.~\cite{GL SU(2)} involves no WZW-term.

The paradox is, however, easily resolved. It
is true in general that a gauge theory is anomaly free if  
the generators $t^i$ of the symmetry group 
satisfy the condition \cite{Gross:1972pv}
\bea
 {\rm Tr} \big(t^i \{ t^j , t^k\}\big) = 0 \co\;\; \forall\;\; i, j,k \fs \non 
\eea
A well known example of such a group is SU(2) -- the above equation obviously
holds if the $t^i$ are identified with the Pauli matrices. Accordingly, in QCD,
the currents of SU(2)$_\indR\times$SU(2)$_\indL$ are anomaly free. 
In ref.~\cite{GL SU(2)}, the investigation
covered only these -- as a consequence
a WZW-term did not appear. The point is that this investigation does not cover
the electromagnetic 
interaction: In the case of two quark flavours, the quark charge matrix
\bea
Q=\diag \{2/3, -1/3\}
\eea
does not represent a generator of the group
SU(2)$_\indR\times$SU(2)$_\indL$ 
because its trace is different from zero. In order to analyze the Ward
identities for 
Green functions that contain the electromagnetic current, we
need to extend 
the symmetry group to SU(2)$_\indR\times$SU(2)$_\indL\times$U(1)$_\indV$. In this 
case, the set of group generators includes the unit matrix and thus the
anomalies fail to vanish.

In the present paper, we analyze the anomalies of the full chiral
group 
U(2)$_\indR\times$U(2)$_\indL$, an investigation which covers the electromagnetic
interaction as well as the anomalies of
the singlet axial current \cite{U(1),KL LargeNCHPT}. 
In section \ref{effective action} (appendix \ref{appdeltaSeff}) we derive the
Ward 
identities of two-flavour QCD and discuss in particular the anomalous
contributions. In the remainder of the paper, we translate the properties of
the underlying theory to the effective language. In section \ref{Low
  energy expansion} we give the leading and next-to-leading order
contributions the the nonanomalous part of the
chiral  
Lagrangian. 
The Wess-Zumino-Witten Lagrangian which accounts for the anomalies in the
context 
of the 
effective theory is considered in 
section \ref{secWZW} (appendix \ref{constrWZW}). For two flavours, this term
is substantially simpler than the general expression and may be written down in
closed 
form. Our findings are illustrated by means of 
two examples of three-point functions in section
\ref{3pt}.

We point out that although, to our best knowledge, the explicit expression for
the 
WZW-term derived in this paper
represents a new result, some of the processes it describes have been studied
earlier: The first low energy theorems for anomalous pion
interactions were derived prior to the 
effective Lagrangian formulation \cite{Steinberger:1949wx}. Furthermore, there
exist a number of publications investigating the 
reactions $ \pi^0 \to \gamma \gamma $ 
\cite{Donoghue:1985wv} or $\gamma(\gamma) \to 3\pi $
\cite{Bijnens:1990ff} 
in 
the context of chiral  
perturbation theory. In these, use was made of the fact that the
effective vertices that describe  
the anomalous interactions of the pions with photons are also contained in the
well  
known expression 
for the WZW-term for SU(3)$_\indR\times$SU(3)$_\indL$. This term itself 
has been the subject of detailed investigations \cite{Issler:1990et}. Its
applications range 
from 
the anomalous decays such as $ \eta \to \pi \pi 
\gamma \gamma$ \cite{Knochlein:1996ah} to weak interaction kaon physics, where
anomalous  
contributions are the rule 
rather than the exception \cite{Bijnens:1992ky}. 

\section{Ward identities}\label{effective action}

We consider quantum chromodynamics with two light flavours in the presence of
external 
vector, axial 
vector, scalar and pseudoscalar fields \cite{GL SU(2)},
\bea \label{LQCD} {\cal L}_{\QCD}\al =\al -\frac{1}{2 g^2}\mbox{tr}
\hspace{-0.5em}\rule[-0.5em]{0em}{0em}_c\hspace{0.4em}
G_{\mu\nu}G^{\mu\nu}  -\theta\, \omega  + \qbar D q + \ldots   
\\
D \al = \al \gamma^\mu (i \partial_\mu + G_\mu + v_\mu+\gamma_5 a_\mu)- s+
i\gamma_5 p \co \non 
\eea 
where the coupling constant $g$ is absorbed in the gluon field $G_\mu(x)$.
The external fields $v_\mu(x)$, $a_\mu(x)$, $s(x)$, $p(x)$ 
represent 
hermitean, colour neutral $2 \times 2 $
matrices in flavour space. The mass matrix of the two light quarks
is contained in the scalar external field
$s(x)$. The vacuum angle $\theta(x)$ represents the variable conjugate
to the operator $\omega(x)$,
\bea 
\omega=\frac{1}{16\pi^2}\,\mbox{tr}
\hspace{-0.5em}\rule[-0.5em]{0em}{0em}_c\hspace{0.4em}
G_{\mu\nu}\tilde{G}^{\mu\nu}
\fs \non \eea
In Euclidean space, the integral
$\nu=\int\! dx\,\omega $ 
is the winding number of the gluon field, so that $\omega(x)$ may be
viewed as the winding number density. The ellipsis in eq.~(\ref{LQCD}) stands
for terms containing the heavy quark fields -- what
matters in the present 
context 
is merely that these
are invariant under the subgroup of chiral
transformations acting in the space of the light flavours,
\bea \label{chitrafo}
q(x) \rightarrow \{V_\indR(x)  \haf (1+\gamma_5 )  + V_\indL(x) \haf
(1-\gamma_5 )\} q(x)\;
\co\;\;\; q = \left( \!\begin{array}{c} u\\d \end{array}\!\right) \co
\eea 
with $V_{\indR}(x), 
V_{\indL}(x)\in\mbox{U(2)}$.

The Green functions of the quark currents and the winding number density are
generated by the effective action, 
\bea 
e^{iS_\eff\{v,a,s,p,\theta\}}=\langle 0\,\mbox{out}|
0\,\mbox{in}\rangle\rule[-0.3em]{0em}{0em}_{v,a,s,p,\theta}\fs \non \eea 
The Ward identities reflect the 
transformation properties of the effective action under
infinitesimal chiral rotations. Formally, the theory
is invariant under the 
transformation in eq.~(\ref{chitrafo}), provided one transforms the external
fields according to 
\bea 
v_\mu' +a_\mu' \al=\al  V_{\indR}  (v_\mu +a_\mu)
{V_\indR}^{\!\dagger} + i V_{\indR} 
\partial_\mu 
{V_\indR}^{\!\dagger}\co \non \\
v_\mu' -a_\mu' \al=\al V_{\indL} (v_\mu -a_\mu) {V_\indL}^{\!\dagger} + i
V_{\indL} 
\partial_\mu 
{V_\indL}^{\!\dagger}\co \non\\
s'+ip'\al=\al
V_{\indR}(s+ip){V_\indL}^{\!\dagger}\fs \non\eea
It is well known that the quark loops spoil this formal symmetry: The
determinant of 
the Dirac operator, $\det D$, is not invariant under the 
transformation specified in eq.~(\ref{chitrafo}). The determinant requires 
renormalization -- there are anomalies because a
regularization that preserves  
chiral symmetry does not exist. 

The change in the Dirac
determinant 
generated by an infinitesimal chiral transformation
\bea \label{alphaRL} V_{\indR}(x)=1+i\?\alpha(x) + i\? \beta(x)\co\hspace{2em} 
V_{\indL}(x)=1+i\?\alpha(x) -i\? \beta(x) \co \non \eea
may be given explicitly.
The modulus of the determinant is invariant 
while its phase picks up two distinct contributions. The
first is proportional 
to the 
winding number density $\omega$. It may be absorbed by the transformation
\bea \label{thetaprim} \theta  \rightarrow \theta -2 \langle \beta \rangle
\eea 
of the vacuum angle (as usual, $\br \ldots \ke$ stands
for the 
trace). This part of the change gives rise to the
U(1)$_\indA$-anomaly and is independent of the presence of external
fields. In contradistinction, the remaining contribution exclusively involves the
external fields and may thus be pulled 
outside the path integral, so that an explicit expression for
the 
change in the effective action can be given. The relevant formula, 
valid for an
arbitrary number of flavours, was
first 
given by Bardeen \cite{Bardeen}.  
We specialize this result to $\nf =2 $, where the explicit expression is 
considerably simpler: With suitably chosen conventions, the
anomalies of two-flavour QCD boil down to\footnote{The sign of $\delta
  S_{\eff}$ 
is convention dependent; we use 
  the metric $\mbox{$+$$-$$-$$-$}$, set 
  $\epsilon_{\sss 0123} = +1$ and identify 
$\gamma_{\sss5}$ with 
  $\gamma_{\sss5} = -i 
  \gamma_{\sss 0}\gamma_{\sss 1}\gamma_{\sss 2}\gamma_{\sss 3}$}
\begin{eqnarray}
\label{delta Seff_2} \delta
S_{\eff}\{v,a,s,p,\theta\}  = 
-\frac{\nc}{16\pi^2} \int \!d^4\!x\, \epsilon^{\mu\nu\rho\sigma}
\br \hat{\beta} 
(\hat{v}_{\mu\nu} +i [ \hat{a}_\mu, \hat{a}_\nu ])\ke  \br
v_{\rho\sigma}\ke \co  
\end{eqnarray}
with $ {v}_{\mu\nu} = \partial_\mu {v}_\nu
-\partial_\nu {v}_\mu  - i [{v}_\mu 
,{v}_\nu] $ and the notation
\bea 
\hat{\beta} = \beta -\haf \br \beta \ke  \co \;\;\; \hat{v}_{\mu\nu} =
v_{\mu\nu} -\haf 
\br  v_{\mu\nu} \ke  \co\;\;\; \hat{a}_\mu = a_\mu -\haf \br a_\mu \ke 
\co \non 
\eea
for the traceless parts of the $2 \times 2 $ matrices $\beta$, $v_{\mu\nu} $
and 
$a_\mu$.  
$\nc$ denotes the number of colours. A derivation of this result on the
basis of Bardeen's formula \cite{Bardeen} may be found in appendix
\ref{appdeltaSeff}. We 
add a few remarks: 

1. The expression in 
eq.~(\ref{delta Seff_2}) is independent of singlet axial vector field $\br
a_\mu  
\ke$. Complications due to the anomalous dimension of the singlet axial current
are thus avoided by the conventions chosen, see appendix
\ref{appdeltaSeff}. 

2. The transformation law in eq.~(\ref{delta
  Seff_2}) states that the effective action is invariant under
  chiral
  U(1)$_\indA$-transformations, where $\beta\propto {\bf 1}$. In view of
  eq.~(\ref{thetaprim}), the anomalies of the
  correlation functions formed with the singlet axial current 
  are thus accounted for collectively by the
  operator 
  identity $(m_q = 0)$
\bea
\partial^\mu\! A_\mu^0 = 2\, \omega \co\;\;\;A_\mu^0 =
\haf\,\qbar 
\gamma_\mu \gamma_5
 q \fs \non 
\eea

3. The remaining anomalies concern correlation functions formed with
the 
currents
\bea
A_\mu^i = \haf\,\qbar \gamma_\mu \gamma_5 \tau^i q\co\;\;\;V_\mu^i = \haf\,
\qbar \gamma_\mu {\tau^i} q \co\;\;\;V_\mu^0 = \haf\, \qbar \gamma_\mu 
q 
 \co \non 
\eea
where the $\tau^i$ denote the Pauli matrices. As eq.~(\ref{delta
  Seff_2}) shows, the vector currents are anomaly
free while for the isovector axial currents $A_\mu^i$ anomalies 
occur in the three 
types of triangle and box 
diagrams shown in the figure: 

\setlength{\unitlength}{1mm} 
\begin{picture}(50,30)
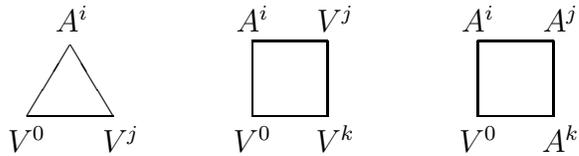
\put(-6.2,20){\parbox[t]{40mm}{Figure 1: Anomalous loop diagrams in
    two-fla\-vour QCD}}
\put(50,10){\line(3,5){5.7}}
\put(50,10){\line(1,0){11.5}}
\put(61.5,10){\line(-3,5){5.7}}
\put(56,23){\makebox(0,0){$A^i$}}
\put(50,7){\makebox(0,0){$V^0$}}
\put(62.5,7){\makebox(0,0){$V^j$}}
\put(80,10){\line(1,0){10}}
\put(80,10){\line(0,1){10}}
\put(80,20){\line(1,0){10}}
\put(90,10){\line(0,1){10}}
\put(80,23){\makebox(0,0){$A^i$}}
\put(91,23){\makebox(0,0){$V^j$}}
\put(80,7){\makebox(0,0){$V^0$}}
\put(91,7){\makebox(0,0){$V^k$}}
\put(110,10){\line(1,0){10}}
\put(110,10){\line(0,1){10}}
\put(110,20){\line(1,0){10}}
\put(120,10){\line(0,1){10}}
\put(110,23){\makebox(0,0){$A^i$}}
\put(121,23){\makebox(0,0){$A^j$}}
\put(110,7){\makebox(0,0){$V^0$}}
\put(121,7){\makebox(0,0){$A^k$}}
\end{picture}

\vspace{-0.7em}\noindent Note that, for two flavours, there are no anomalies
in pentagon diagrams.

4. The expression for the change in the effective action reduces to zero 
if the singlet vector field vanishes, $\br v_\mu 
\ke= 0$, in accordance with the statement that Green functions 
built
exclusively with the currents of SU(2)$_\indR\times$SU(2)$_\indL$ are
anomaly free. Stated in terms of physical degrees of freedom: There are no
anomalies in two-flavour QCD unless at least one of the neutral gauge bosons,
$\gamma$ 
or $Z$, is involved, as in the reactions $\pi^0 \to \gamma \gamma$,
$\pi^-  
\to 
e^-  \bar{\nu}_e \gamma$, $\gamma \pi^0   \to
\pi^+ \pi^- $.

5. As shown by Wess and Zumino \cite{WZ}, the fact that the Ward
identities follow from the variation of a single functional imposes nontrivial
constraints on the structure of the anomalies. The expression (\ref{delta 
  Seff_2}) obeys the relevant 
consistency conditions.

\section{Low energy expansion}\label{Low energy expansion}

In view of the U(1)$_\indA$-anomaly, the symmetry group of the massless theory
is 
SU(2)$_\indR \times$SU(2)$_\indL\times$U(1)$_\indV$.
We consider the standard scenario and assume that this symmetry 
is spontaneously broken, the ground state being
invariant only under the subgroup U(2)$_\indV$.
The low energy properties
of the theory 
are then governed by the pions, the three Goldstone degrees of freedom
associated 
with 
this breakdown \cite{Weinberg Physica,GL SU(2),GL SU(3)}.
It is convenient to collect 
 these fields in a matrix $U(x) \in \mbox{U($2$)}$ that transforms 
according to the representation
\bea U'(x)= {V}_\indR(x) U(x) {{V}_\indL}^{\!\dagger}(x)\fs\eea  
The standard constraint $\det U=1$ is not consistent with 
this transformation law, because the phase of the determinant is not
preserved. We replace it by the condition \cite{GL SU(3)} 
\bea
\det U (x) = e^{-i \theta (x) }\co
\eea
which is in accordance with the transformation properties of
the vacuum angle (\ref{thetaprim}). 

The form of the effective Lagrangian is determined by the symmetry properties
of the underlying theory \cite{Weinberg Physica,GL SU(2),GL
  SU(3),Leutwyler:1994iq}: The low energy expansion of the effective action of
QCD is reproduced if the effective Lagrangian 
consists of (a) the Wess-Zumino-Witten term that reproduces the correct
anomalies (see section \ref{secWZW}) and 
(b) the most general expression consistent with chiral symmetry that can
be built with the fields $U,$ $v_\mu,$ $a_\mu,$ $s,$ $p$ and $\theta$ and
their derivatives. The terms in the effective Lagrangian are ordered according
to their low energy dimension,
\bea
{\cal L}_\eff = {\cal L}^{(2)} +{\cal L}^{(4)} + \ldots  \non
\eea
where the first term is of order $p^2$, the second of order $p^4$,
etc\footnote{We use the standard bookkeeping as introduced in \cite{GL
    SU(2),GL SU(3)}}. In view of  
the $\epsilon$-tensor, the WZW-term only shows up at order $p^4$.
The Lagrangian of order $p^2$ is thus gauge
invariant,   
\bea\label{Leff 2}
{\cal L}^{\sss (2)} = \kwa {F^2}\br D_\mu U^\dagger D^\mu U+
U^\dagger 
\chi + 
\chi^{\dagger} U\ke + \tfrac{1}{8} h_0\? D_\mu \theta D^\mu \theta
\co 
\eea
with $\chi \equiv 2 B (
s+ i p)$ and the covariant derivatives
\bea\label{DU}
D_\mu U = \partial_\mu U- i\? {r}_\mu\? 
U+ i\?
U\?  {l}_\mu +\tfrac{i}{2} D_\mu \theta\? U \co\; \;\; D_\mu \theta =
\partial_\mu 
\theta + 
2 \br a_\mu \ke \co 
\eea
where $r_\mu = v_\mu + a_\mu$, $l_\mu = v_\mu - a_\mu$.  
The covariant derivative $D_\mu U$ transforms in the same manner as $U$,
$D_\mu
\theta$ is gauge invariant. Note that we have defined the covariant derivative
of $U $ 
in such a way that the trace $\br U^\dagger D_\mu U \ke$ vanishes. 

At order $p^4$, the effective Lagrangian receives two categories of
contributions, 
one of which
involves the $\epsilon$-tensor while the other does not. Accordingly we
decompose the $p^4$-Lagrangian in the natural parity part ${\cal L}^{\sss
  (4)}_g$ (which does not involve the $\epsilon$-tensor)
and the unnatural parity part ${\cal L}^{\sss
  (4)}_\epsilon$, 
\bea
{\cal L}^{\sss (4)} ={\cal L}^{\sss (4)}_g  + {\cal L}^{\sss (4)}_\epsilon
\fs \non 
\eea

The natural parity part represents a gauge invariant expression. Using the
equations of motion associated with the Lagrangian in 
eq. (\ref{Leff 2}), the most general expression consistent with chiral
symmetry, 
$C$ and $P$ reads 
\bea \label{Leff 4}
{\cal L}^{\sss (4)}_g \al=\al {\ell_1}\?\tfrac{1}{4}\br D_\mu
U^\dagger
D^\mu U 
\ke   
^2 +{\ell_2}\?\tfrac{1}{4}\br D_\mu U^\dagger D_\nu U \ke \br D^\mu
U^\dagger D^\nu U \ke  
\non\\  
\al\al \hspace{0em}+ {\ell_3}\? \tfrac{1}{16} \br U^\dagger
\chi + 
\chi^\dagger U 
\ke^2\non 
 + \? {\ell_4}\?\tfrac{1}{4} \br D_\mu U^\dagger
D^\mu \chi+  D_\mu  
\chi^\dagger D^\mu U   \ke \non\\
\al\al \hspace{0em} 
 + 
 \ell_5\? \br U^\dagger \hat{R}_{\mu \nu} U \hat{L}^{\mu \nu}\ke  +\?
{\ell_6}\? \tfrac{i}{2} \br \hat{R}_{\mu \nu} D^\mu U D^\nu
U^\dagger + \hat{L}_{\mu \nu} D^\mu U^\dagger  D^\nu U \ke
\non\\  
\al\al\hspace{0em}
- {\ell_7}\?\tfrac{1}{16}\br  U^\dagger \chi -
\chi^\dagger U   
\ke^2  + \ell_{8}\? D_\mu \theta D^\mu \theta \br D_\nu
U^\dagger  
   D^\nu U \ke \non\\ 
\al\al\hspace{0em}+ \ell_{9}\? D_\mu \theta D_\nu  \theta \br D^\mu
   U^\dagger D^\nu U\ke
 + \ell_{10} \? D_\mu \theta D^\mu \theta \br
   U^\dagger  \chi  +\chi^\dagger U \ke \\ \al\al \hspace{0em}
 -\?
   \ell_{11}\?i\?  D_\mu 
   \theta \br D^\mu U^\dagger \chi-  D^\mu U \chi^\dagger
   \ke  
 +  \ell_{12}\? i\? \partial_\mu D^\mu \theta \? \br
   U^\dagger \chi - \chi^\dagger U \ke  \non \\ 
\al\al\hspace{0em} +\?
\tfrac{1}{4} (h_1 + h_3 ) \br 
\chi^\dagger \chi \ke\non   
+ \tfrac{1}{4} (h_1 - 
h_3 ) ( \det \chi^\dagger e^{-i\theta} + \det \chi\, e^{i\theta} ) \non \\
\al\al\hspace{0em}  -\? \tfrac{1}{2}  ( \ell_5+ 4 h_2 ) \br
\hat{R}_{\mu \nu} \hat{R}^{\mu \nu}+ 
\hat{L}_{\mu \nu}\hat{L}^{\mu  \nu}\ke
    + h_4\? \tfrac{1}{4}  \br
   R_{\mu\nu}+L_{\mu\nu}\ke\br R^{\mu\nu}  +L^{\mu\nu}\ke \non\\ 
\al\al\hspace{0em} +\?  h_5\?  
   \tfrac{1}{4}\br 
   R_{\mu\nu}-L_{\mu\nu}\ke\br R^{\mu\nu}-L^{\mu\nu}\ke  
+ h_6\? ( D_\mu \theta D^\mu \theta )^2 + h_7 \?( \partial_\mu
   D^\mu 
   \theta )^2  \co \non
\eea
with the right- and lefthanded field strengths
\bea \al\al  R_{\mu\nu} = \partial_\mu r_\nu - \partial_\nu r_\mu - i [r_\mu,
r_\nu] 
\co \hspace{2em}\hat{R}_{\mu\nu} = {R}_{\mu\nu}- \haf \br { R_{\mu\nu}} \ke
\co \non \\  
\al\al L_{\mu\nu} = \partial_\mu l_\nu - \partial_\nu l_\mu - i [l_\mu, l_\nu]
\co \hspace{2.7em} \hat{L}_{\mu\nu} = {L}_{\mu\nu}- \haf \br { L_{\mu\nu}} \ke
\fs  \non 
\eea
The vertices proportional to $\ell_1,\ldots\,,\ell_7$ and
$h_1,h_2,h_3$ are those relevant for the low energy expansion of the Green
functions formed with the 
SU(2)$_\indR\times$SU(2)$_\indL$ currents. This part of 
the effective Lagrangian was given by Gasser and
Leutwyler \cite{GL SU(2)} and we have taken over
their notation\footnote{In ref.~\cite{GL SU(2)}, the effective
  field is 
  characterized by a four component vector $U^\indA$.
The
  relation between the two representations reads: $U^0 = \frac{1}{4}\br
  U^\dagger  
  + U \ke$, $U^k = \frac{i}{4}\br \tau^k (U^\dagger -U) \ke $}.
The additional terms proportional to $\ell_8,\ldots,\ell_{12}$ and $
  h_4,\ldots,h_7$ 
are needed for the study of correlation functions involving the operators
  $A_\mu^0$, 
  $\omega$ or $V_\mu^0$.

The
unnatural parity part involves a single free parameter at order $p^4$,
\bea \label{Lunnp}
{\cal L}^{\sss (4)}_\epsilon = {\cal L}_\WZW+
i\?\tilde{\ell_{1}}\?  
   \epsilon^{\mu\nu\rho\sigma}  D_\mu \theta \br \hat{R}_{\nu\rho} D_\sigma 
   U U^\dagger -  \hat{L}_{\nu\rho} U^\dagger D_\sigma U
   \ke \fs
\eea
In the absence of singlet fields this part of the Lagrangian vanishes
altogether. The explicit expression for the Wess-Zumino-Witten term ${\cal
  L}_\WZW$ will be 
given in the 
following  
section. 

In appendix B of ref.~\cite{KL LargeNCHPT}, the extensions required by the
presence of the singlet fields $\br v_\mu \ke$, $\br a_\mu \ke$ and 
$\theta$ were discussed in the framework of the effective theory with three
light 
flavours. We briefly mention two observations made there that also apply in the
present case:

1. It is well known that the dimension of the singlet axial current, $A_\mu^0$,
is anomalous \cite{Kodaira:1980pa}. The implications of this fact for the
effective theory were worked
out in refs.~\cite{KL LargeNCHPT,Leutwyler:1998yr}: In the effective
Lagrangian, the 
field $\br a_\mu \ke$ as well as some of the effective coupling constants
depend on the running scale of QCD. 
In the present case, the only coupling
constants that do depend on the scale are those which multiply terms that
involve the covariant derivative $ D_\mu \theta$ or the field strength of the
singlet axial current, $\br 
   R_{\mu\nu}-L_{\mu\nu}\ke$. Under a change of scale these undergo
   multiplicative renormalization, while all other fields in the effective
   Lagrangian stay put -- in particular, the definition of the
   covariant derivative 
   $D_\mu U$ in eq.~(\ref{DU}) was chosen on this purpose. The renormalization
   group invariance of the effective Lagrangian is ensured by the requirement
   that the effective coupling constants transform contragrediently to the
   terms they multiply. In particular, the coupling constants $F, \ell_1,
   \ldots, \ell_7$ of ref.~\cite{GL SU(2)} do not depend on the running
   scale of QCD. The scale dependence of the coupling constants $ \ell_{11} $
   and 
   $\ell_{12}$, for instance, reflects the fact that the matrix element $\br
   0| A_\mu^0 | \pi^0 \ke$ does not represent a measurable quantity.

2. The second remark concerns the renormalization within the effective
theory -- note that this issue is completely unrelated to the one discussed
above. The 
one loop graphs  
associated with the Lagrangian given in eq. (\ref{Leff 2}) require
renormalization. In ref.~\cite{GL
  SU(2)}, the relevant counterterms have been worked out in the absence of the
singlet  
fields $\br v_\mu \ke$, $a_\mu$ and $\theta$. It was shown that
the infinities may be absorbed in the coupling 
constants  
$\ell_1,\ldots,\ell_{6}$ and $ h_1,h_2$. As discussed in 
ref.~\cite{KL LargeNCHPT}, this situation is not altered by the presence of the
singlet fields: None of the coupling constants $\ell_8, \ldots, \ell_{12}$,
$h_4, 
\ldots, h_7$ or $\tilde{\ell}_1$ is needed to renormalize the one loop graphs
of the
effective theory. 

\section{Wess-Zumino-Witten term}\label{secWZW}

So far, we have discussed the gauge invariant contributions to the effective
Lagrangian. In view of the fact that the 
effective theory does not involve fermionic degrees of freedom, 
the effective action generated by a gauge invariant Lagrangian
represents a gauge invariant object -- the anomalies of the
underlying theory are not accounted for. Even before effective Lagrangians
were studied systematically, Wess and Zumino \cite{WZ} pointed out how the
anomalies can be accounted for in this framework: In
addition to the gauge 
invariant Lagrangian one introduces a gauge variant functional of the
effective fields,
$S_\WZW$, constructed 
so as to reproduce the change in the effective action
\cite{WZ,GL SU(3),W,WZW,KT}. Clearly, the 
crucial 
point here is that the functional is allowed to depend
on the Goldstone field $U$,
$S_\WZW =S_\WZW \{U,v,a,\ldots\}$ -- the
very existence of anomalies is a manifestation of the
statement that a local expression formed with the
external fields alone cannot generate the change in (\ref{delta Seff_2}). 
In this way, the anomalies of the
underlying 
theory generate effective vertices for the 
Goldstone fields. 

In the present case, the Wess-Zumino-Witten term does not involve the singlet
axial field, $S_\WZW =S_\WZW \{U,
v,\hat{a}\}$. It is given by the action of a suitable
Lagrangian, $S_\WZW = \int \!d^4\!x\, {\cal
  L}_\WZW$, which can be written down in closed form:
\bea\label{LWZW}
\al\al\hspace{-2em}{\cal L}_\WZW = - \frac{\nc} {32 \pi^2 }\,
\epsilon^{\mu\nu\rho\sigma} \big\{ \br 
U^\dagger \hat{r}_{\!\?\mu} \? U\? 
\hat{l}_\nu - \hat{r}_{\!\?\mu} \? \hat{l}_\nu + i{\Sigma}_\mu
(U^\dagger \hat{r}_{\!\?\nu} \?
U   + \hat{l}_\nu) \ke \br v_{\rho\sigma} \ke \\ 
\al\al \hspace{7.65em} + \tfrac{2}{3}\br
{\Sigma}_\mu {\Sigma}_\nu {\Sigma}_\rho \ke \br v_\sigma \ke
\big\} 
\co\non\eea
where the quantities $\hat{r}_\mu,\hat{l}_\mu$ and ${\Sigma}_\mu$ stand for
\bea \hat{r}_\mu = \hat{v}_\mu + \hat{a}_\mu\co \;\;\;\; \hat{l}_\mu =
\hat{v}_\mu - \hat{a}_\mu 
\co \;\;\;\; {\Sigma}_\mu =U^\dagger\partial_\mu U  \fs
\non 
\eea 
A derivation of this result is presented in appendix~\ref{constrWZW}.
To prove that this expression accounts for the anomalies of two-flavour QCD
correctly, it suffices, however, to show that, under an infinitesimal chiral
rotation of 
the fields $U$, $v$ and $\hat{a}$, it does generate the change 
given in
eq.~(\ref{delta Seff_2}).
That this is indeed the case may be verified by explicit
calculation.
It may also be checked that the expression given in eq.~(\ref{LWZW})
respects parity and charge 
conjugation invariance. Naturally, the Wess-Zumino-Witten term is unique only
up to gauge invariant contributions. Here, we have adapted the convention $
S_\WZW\{{\bf 1}, v,\hat{a} \}  = 0$ (see
appendix~\ref{constrWZW}).  

As anticipated 
in the introduction, the Wess-Zumino-Witten Lagrangian in
eq.~(\ref{LWZW}) is considerably simpler than the standard expression
for this term \cite{W,WZW,KT}. In particular,
one notices that the term 
identified  
by Witten \cite{W}, the five-dimensional
integral of the form  
\bea \non 
\int_{M_5} \!d^5\!x\, \epsilon^{ijklm} \langle
{\Sigma}_i{\Sigma}_j{\Sigma}_k{\Sigma}_l{\Sigma}_m \rangle 
\eea
is missing. The absence of this term is due to the fact that
there is no anomaly in the correlation function of five axial vector
currents. Its absence can also be understood on the level of the effective
theory: 
The four 
available
fields, $\pi^+,\pi^-,\pi^0$ and $\theta$, do not allow for a nonvanishing,
totally antisymmetric fifth rank tensor. This implies that the  
five-point function is not only free from anomalies but, moreover, vanishes to
the order 
considered. 
For two flavours, the anomalous interactions of the Goldstone fields are
absent if the 
external fields are switched off. In fact, eq.~(\ref{LWZW}) shows that they
are proportional to the trace 
$\br v_\mu \ke$ of the vector field, as was already discussed in
section~\ref{effective action}.

We add two remarks in connection with the U(1)$_\indA$-anomaly. 
First, we note that the Lagrangian (\ref{LWZW}) is independent of the
singlet  
axial field $\br a_\mu \ke$. Accordingly, it represents a renormalization
group invariant  
expression. Furthermore, this expression is independent of the vacuum angle
$\theta(x)$. 
This follows from the fact that the Lagrangian in eq.~(\ref{LWZW}) is
invariant under chiral U(1)$_\indA$-rotations: The Lagrangian is the same
independently of whether the matrix $U$ is subject to the constraint $\det U =
e^{-i \theta} $ or
$\det U = 1$.

\section{Three-point functions}\label{3pt}

As an illustration, we give here the expressions for two examples of
three-point  
functions determined by the unnatural parity part of the Lagrangian
(\ref{Lunnp}). As the first example we consider the correlation function of
one axial and two vector currents, $
\br 0 | {\rm T} A^i V^k V^0 |0 \ke$, which, to the order considered, is
completely determined by the Wess-Zumino-Witten term. This correlation function
determines the radiative decay of the neutral pion and contributes also to
$ \pi^- \to e^- \bar{\nu}_e \gamma $. The term relevant for the present
calculation is
the 
second one in the expression for the WZW-Lagrangian
(\ref{LWZW}). The third contributes in an analogous manner, for instance,
to the process
$\gamma\pi^0  \to  \pi^+\pi^- $, while the first only matters when at least
three gauge   
bosons are involved. A simple calculation leads to the result
\bea \hspace{-1 em} i^2 \!\!\int \!\!dx \,dy\, e^{-i kx+ip y +iq
  z}\br 0 | 
{\rm T} 
  A^{i}_{\alpha}(x) V^{k}_{\mu}(y) V^{0}_{\nu}(z) | 0 \ke \non \al \al \\
 \hspace{-8em}
  \al \al \hspace{-8em} = \;  i\frac{\nc }{8 \pi^2
  }\,\frac{\;\;\delta^{ik}}{{M_i}^{\!2}-k^2}\,   
   \epsilon_{\mu \nu \rho \sigma}\, k_\alpha\, p^\rho q^\sigma + O (p^3) \co
  \non  
\eea
with $k =p+q$, $M_1=M_2 = M_{\pi^+}$, $M_3 = M_{\pi^0}$. Upon contracting the
result with $k^\alpha$, the anomalous divergence of this correlation function
is easily 
identified (note that, for nonvanishing quark masses, the Ward identity
involves 
the correlation function $\br 0 | {\rm T} P^i V^k V^0 |0 \ke$ of the
pseudoscalar current $P^i =
\qbar i \gamma_5 \tau^i q$).
 The residue at 
$k^2 
={M_{\pi^0}}^{\!\!\!\!\!\!2}\,\,$ determines the $ \pi^0 \to \gamma
\gamma 
$ amplitude\footnote{
The electromagnetic current $J^{\,\mbox{\scriptsize{e.m.}}}_\mu$
also receives a contribution from the 'heavy' quark
  flavours $s$, $c$, $b$, $t$,  
\bea \non
J^{\,\mbox{\scriptsize{e.m.}}}_\mu = J_\mu^{\ell} +J_\mu^h \co \;\;\;\;
J^h_\mu
= \tfrac{2}{3} \bar{c} \gamma_\mu c - \tfrac{1}{3} \bar{s} 
\gamma_\mu s+\tfrac{2}{3} \bar{t}\gamma_\mu t - \tfrac{1}{3} \bar{b}
\gamma_\mu b 
\fs \eea 
To account for the contributions from the heavy quarks, we extend the list of
external fields in eq.~(\ref{LQCD}) so as to include a source field $A^\mu$
for the current $J_\mu^h$
\bea \non
{\cal L}_{\QCD} \rightarrow {\cal L}_{\QCD} - e A^\mu J_\mu^h\fs
\eea
Note that the current $J_\mu^h$ is conserved. The extended framework thus
exhibits an additional local U(1)$_\indV$-symmetry: The effective action is
invariant under gauge transformations 
of $A^\mu$. Starting at 
order $p^4$, gauge symmetry permits corresponding additional terms in
the effective theory:
\bea \non
{\cal L}_\eff \rightarrow {\cal L}_\eff + c_1 \br v_{\mu\nu} \ke
F^{\mu\nu}  + c_2 F_{\mu\nu} F^{\mu\nu}  + O(p^6) \co
\eea 
with $F_{\mu\nu} = \partial_\mu A_\nu - \partial_\nu A_\mu$. The new terms,
however, generate only contact contributions. To the order considered and what
concerns matrix elements of 
pions, the electromagnetic current may therefore be identified with
$J_\mu^{\ell}= 
\tfrac{2}{3} \bar{u} \gamma_\mu u - \tfrac{1}{3} \bar{d}  
\gamma_\mu d$}. Integration over phase space leads to the well known 
prediction
\bea\non 
\Gamma_{\pi^0 \to \gamma \gamma }  = \frac{\alpha^2 \nc^2 
  {M_{\pi^0}}^{\!\!\!\!\!\!3}}{576 \,\pi^3 {F_\pi}^{\!\!2}}+
O(m^\frac{5}{2})\co 
\eea
where $\alpha $
  denotes the fine structure constant. 

In contrast to the above example, the correlation function $\br 0 | {\rm T} A^i
A^k A^0 |0 \ke $ does not receive a contribution from the WZW-term but is
instead 
proportional to the effective coupling constant $ \tilde{\ell}_1
$,  
\bea
\al \al  \hspace{-1em} i^2 \!\!\int \!\!dx \,dy\, e^{i px+iq y -ik
  z}\br 0 | 
{\rm T} 
  A^{i}_{\mu}(x) A^{k}_{\nu}(y) A^{0}_{\rho}(z) | 0 \ke
  =  T^{ik}_{\mu\nu\rho} (p,q,k ) +  T^{ki}_{\nu\mu\rho} (q,p,k )\co \non \\
\al\al \hspace{2em} T^{ik}_{\mu\nu\rho}(p,q,p+q ) =
 8\? i\? {\tilde{\ell}_1}\,\delta^{ik} 
 \epsilon_{\nu \rho\alpha \beta}\,  q^\alpha
 \Big\{ \frac{\;\;p_\mu p^\beta}{{{M_i}^{\!2}- p^2 }}  + g_{\mu}^{\;\;\beta}
 \Big\} + O (p^3) \fs 
  \non 
\eea
In this case the divergences of the charged
axial 
currents are anomaly free. In the present framework, this is
true 
in general for 
the correlation 
functions involving the singlet axial current since these only
receive contributions from the gauge invariant part of the Lagrangian -- the
WZW-term is independent of the field $\br a_\mu \ke$. 
In the effective Lagrangian, the anomalies of the singlet axial current
manifest themselves in that the the field $\br a_\mu \ke$ exclusively
enters in 
combination with the vacuum angle, in the form $ 
D_\mu \theta = \partial_\mu \theta + 2 \br a_\mu  \ke $.
One may check that in the above example the 
divergence of the singlet axial current is indeed proportional to correlation
function  
$\br 0 | {\rm T} A^i  
A^k \omega |0 \ke $.

\section*{Acknowledgments}

I am grateful to H. Leutwyler for many illuminating discussions concerning the
subjects of the 
present paper. I also thank  
J. Gasser for useful remarks. Furthermore, I profited from discussions with
T. Becher, P. Liniger and J. Schweizer.

\begin{appendix}

\section{Anomalies}\label{appdeltaSeff}

In the present appendix we derive the expression for the anomalies in
two-flavour QCD given in eq.~(\ref{delta Seff_2}). For
the 
following considerations it is convenient to introduce the
differential 
forms 
\bea v\al=\al dx^\mu v_\mu\co\hspace{2em}
a=  dx^\mu a_\mu\co\hspace{2em} 
d= dx^\mu\partial_\mu\fs
\nonumber\eea 
The quantities $dx^0,dx^1,dx^2,dx^3$ are treated as Grassmann variables. 
Their product yields the standard volume element, 
$dx^\mu dx^\nu dx^\rho dx^\sigma=\epsilon^{\mu\nu\rho\sigma}d^4\!x$.

The effective action of QCD is defined only up to contact terms formed with
the external fields. It may be chosen such that it is
invariant under the 
transformation generated by the vector charges. Accordingly, the change of the
effective action under
an infinitesimal chiral rotation (\ref{alphaRL}) may be written in the form
\bea \label{deltaSold} \delta
S_{\eff}\{v,a,s,p,\theta\} = 
-  \int 
\br \beta \,
\Omega(v,a) \ke \fs
\eea
An explicit formula for $\Omega(v,a)$ was first given by
Bardeen  
\cite{Bardeen} -- it is valid for an arbitrary number of flavours $\nf$:
\bea  
 \al\al\Omega(v,a) =
\frac{\nc}{4\pi^2}\left\{\Fv\Fv +\mbox{$\frac{1}{3}$}\Dva \Dva
+\mbox{$\frac{i}{3}$} (\Fv {a}^2+4 {a}
  \Fv {a} + {a}^2 \Fv)+\mbox{$\frac{1}{3}$} {a}^4\right\}\co \non \\ 
\al\al \Fv= dv-iv^2\co
\hspace{2em}\Dva =d{a}-iv{a}-i{a}v\co\non\eea
where $v$ and $a$ are $\nf\times\nf$ matrices. Decomposing these
into their traceless parts $\hat{v}$, $\hat{a}$  
and a remainder,
\bea
v = \hat{v} + \frac{1}{\nf} \br v \ke \co \hspace{3em} a = \hat{a} +
\frac{1}{\nf} \br a 
\ke  \co \non  
\eea  
$\Omega $ may be written as a sum of three terms, $ \Omega = \Omega_1 +
\Omega_2 + \Omega_3$, 
\bea \label{Omega123}
\Omega_1 &=& \Omega(\hat{v},\hat{a}) - \frac{1}{\nf} \br
\Omega(\hat{v},\hat{a}) \ke 
\co \\
\Omega_2 &=&  \frac{\nc}{2 \nf \pi^2 } \{ \hat{F}_v + i \hat{a}^2 \} \br dv
\ke 
\co  \non\\
\Omega_3 &=&  \frac{\nc}{6 \nf \pi^2 } \{ \Dvahut \br da \ke  -2
i(\Fvhut   \hat{a}- \hat{a} \Fvhut ) \br a \ke    \} + \frac{1}{\nf} \br
\Omega(v,a) \ke \co  \non
\eea with $\Fvhut = d\hat{v}-i\hat{v}^2$,
$\Dvahut 
=d\hat{a}-i\hat{v}\hat{a}-i\hat{a}\hat{v}$. The significance of this
decomposition is the following: The first term, $\Omega_1$, is the anomaly
corresponding to the well known standard case where the singlet vector and
axial vector 
currents 
are disregarded and the chiral transformations are
restricted to the subspace SU($\nf$)$_\indR\times$SU($\nf$)$_\indL$. For
$\nf=2$, the
second term, $\Omega_2$, is identical with the expression for the anomaly
given 
in 
eq.~(\ref{delta Seff_2}) -- its implications are discussed in detail in the
present 
paper. The third term, 
$\Omega_3$, has the property that it represents
the anomaly  
generated by a contact  
term $P$,
\bea
\;\; P  = \frac{\nc} {8   \nf \pi^2 }  \big\{ \br v (d v -\tfrac{2i}{3}
  v^2  
  ) \ke d
\theta + \br a D_v a \ke ( \tfrac{4}{3} \br a \ke + d \theta)  \big\} -
\frac{\nc} {12   \nf^2 \pi^2 } \br a \ke \br da \ke d \theta \co
\non  
\eea 
with $\delta \!\int\!  P = - \br \beta\, \Omega_3 \ke $.
As stated above, we may redefine the effective action by removing this
term, 
\bea
S_\WZW \rightarrow S_\WZW - \int \! P \fs \non 
\eea 
With these conventions, the anomalies of $\nf$-flavour QCD are given by 
\bea \delta \label{deltaSnew}
S_{\eff}\{v,a,s,p,\theta\} = 
-  \int \br \hat{\beta} \,(\Omega_1 + \Omega_2 ) \ke \co 
\eea
where we used the fact that both, $\Omega_1$ and $ \Omega_2$, are
traceless. For two flavours, $\nf = 2$, the term $\br \hat{\beta} \,
\Omega_1 \ke $
reduces to zero and we 
are left with the result given in eq.~(\ref{delta Seff_2}). 

Above, we have made use of the freedom in the definition of
the 
effective action to 
bring the anomalies to a simple form. Correlation functions
calculated with either version of the effective action
differ only in contact 
contributions which are physically irrelevant. In one respect, however, our
definition of the anomalies is conceptually 
superior to the 
original one: As 
pointed out in ref.~\cite{KL LargeNCHPT}, the form of the anomaly in
eq.~(\ref{deltaSold}) fails to be renormalization group invariant, 
because it involves the singlet axial field $\br a_\mu \ke$. As a consequence
of the anomalous dimension of the singlet current $A^0_\mu$, this field 
depends on the running scale of QCD -- otherwise the perturbation generated by
the term $\langle a_\mu\rangle A^\mu_0$ would fail to be renormalization group
invariant. While the right hand side of eq.~(\ref{deltaSold}) is in conflict
with the scale independence of the theory, eq.~(\ref{deltaSnew}) 
does represent a renormalization group
invariant statement. 

The requirement of renormalization group invariance only constrains that 
part of contact term $P$ which involves the field $\br a_\mu \ke$. In
addition, 
we have adapted here the convention that the change is independent of $\br
\beta \ke$ and $\theta$, as a consequence of which the 
  anomalies of the 
  singlet axial current become universal.

\section{Construction of the WZW-term}\label{constrWZW}

In this appendix, we demonstrate how the Wess-Zumino-Witten term may be
obtained from the explicit expression for the change in the effective action
given in eqs. (\ref{delta Seff_2}) and (\ref{Omega123}) \cite{WZ,GL
  SU(3),W,WZW,KT}. We seek a functional $S_\WZW   
  \{{U}, v, \hat{a}  \} $ with the property that it reproduces the anomaly
  under an 
infinitesimal chiral transformation,  
\bea \label{diffeq}
D(\alpha, \beta) S_\WZW   
  \{{U}, v, 
\hat{a}  \}   = - \int \br \hat{\beta}\, \Omega_2 \ke \co  \eea  
where $D(\alpha, \beta)$ denotes the generator of infinitesimal chiral
  transformations \cite{WZ,GL SU(3)}. The exponential of $D(\alpha, \beta)$
  generates finite chiral transformations: 
\bea \label{gaugetrs}
e^{D(\alpha, \beta)} U e^{-D(\alpha, \beta)} \al =  \al
  V_\indR  U 
  {V_\indL}^{\!\dagger}  \co \non \\ 
  e^{D(\alpha, \beta)} (v+a )   e^{-D(\alpha, \beta)} \al = \al V_\indR
  (v+a){V_\indR}^{\!\dagger}   + i V_\indR  d {V_\indR}^{\!\dagger} \co \\
  \non 
e^{D(\alpha, \beta)} (v-a)    e^{-D(\alpha, \beta)} \al =  \al V_\indL
  (v-a)   {V_\indL}^{\!\dagger} 
   + i V_\indL  d {V_\indL}^{\!\dagger} \co 
\eea 
with $V_\indR = e^{i (\alpha +\beta) }$, $ V_\indL = e^{i (\alpha -
  \beta) }$. In order to
render its  
solution unique, we proceed to
equip the differential equation  
(\ref{diffeq}) with a boundary condition,
\bea \label{boundaryc}
S_\WZW   
  \{{\bf 1}, v, 
\hat{a}  \} = 0 \fs 
\eea
Note the consistency of this equation with the invariance with
respect to vector 
transformations.  
(Eq.~(\ref{diffeq}) shows that 
the Wess-Zumino-Witten functional is invariant 
under U(1)$_\indA$-rotations,
$S_\WZW    
  \{{U}e^{i \varphi}, v, 
\hat{a}  \} =S_\WZW   
  \{{U}, v, 
\hat{a}  \}$. Here and in the following we may therefore disregard the
associated 
ambiguities -- the final result will be independent of these.)
From the two preceding equations one infers 
\bea \non 
e^{D ( \alpha , \beta )} S_\WZW   
  \{{U} , v, 
\hat{a}  \}\,\rule[-0.5em]{0.04em}{1.4em}_{\;\alpha=\alpha_\indU,\,
\beta=\beta_\indU} = 0
\eea 
for a pair of matrices $\{ \alpha_\indU,\beta_\indU \}$ that satisfy the
equation  
\bea 
e^{i({\alpha_\indU}+ {\beta_\indU}) } {U} 
e^{- i({\alpha_\indU}-{\beta_\indU}) } =
  {\bf 1} \fs  \non 
\eea
The solution
  to this equation is obviously not unique -- the Wess-Zumino consistency
  conditions \cite{WZ} guarantee, however, that the final result will be
  independent of the 
  choice\footnote{In fact, at the cost of some algebraic complication, the 
    subsequent steps may also be carried out without specifying a particular
    solution of the type (\ref{choice})}. A
  particularly  
  convenient 
choice is given by \cite{KT} 
\bea \label{choice} \alpha_\indU = - \beta_\indU \co \;\;\; 
U = e^{-2i \beta_\indU}  \co \;\;\;  {\rm i.e.}\;\;\;  V_\indR
  ={\bf 1}\co \;\;\; V_\indL  = {U} \fs 
  \eea 
Putting things together, we arrive at  
\bea \label{SWZW}
S_\WZW   
  \{{U} , v, 
\hat{a} \} \al = \al  \sum_{n=1}^\infty \frac{1}{n!}\int  \br
\hat{\beta}\, [D
  (\alpha, \beta)]^{n-1}   
\Omega_2 \ke\,\rule[-0.5em]{0.04em}{1.4em}_{\;\alpha=\alpha_\indU,\,
\beta=\beta_\indU}  \non \\
\al = \al  \int_0^1 dt
  \int \br 
   \hat{\beta}\, e^{D
   (t \alpha  ,t \beta  )}   
\Omega_2 \ke\,\rule[-0.5em]{0.04em}{1.4em}_{\;\alpha=\alpha_\indU,\,
\beta=\beta_\indU}
\eea
where we used the linearity of $ D$ with respect to its arguments. 
When acting on $\Omega_2$, the operator
  $ e^D$ replaces the fields $v$ and $ \hat{a} $ with their
  gauge transforms, cf. eq. (\ref{gaugetrs}):   
\bea e^{D
  ( t \alpha  ,t\beta  )} \,  \Omega_2(v, \hat{a} )  = \Omega_2
  (v_t,\hat{a}_t)  \fs \non \eea
Evaluating this at $\alpha=\alpha_\indU,\,\beta=\beta_\indU$ from
eq.~(\ref{choice}), we obtain 
\bea \;\;\;
{v}_t + {a}_t ={v}  + {a} \co \;\;\;\; {v}_t-{a}_t = {U}_t ({v}  -
{a}){{U}_t}^\dagger +i 
  {U}_t d {{U}_t}^\dagger \co \;\;\;\; {U}_t = e^{-2it \beta_\indU }  \fs
  \non 
\eea
Finally, using 
$ \partial_t {U}_t = -2 i \beta_\indU {U}_t$, 
the integration with respect to $t$ in eq. (\ref{SWZW}) can be
performed explicitly. The result is of the form
\bea
S_\WZW   
  \{{U} , v, 
\hat{a} \} = \int \!\! d^4\!x\, {\cal L}_\WZW \co \non
\eea
and, for $\nf =2$, the explicit expression for ${\cal L}_\WZW$
is the one in  
eq.~(\ref{LWZW}).

The above derivation explicitly shows that, apart from 
the effective variables, the WZW-term exclusively involves the fields which
occur also 
in the expression for 
the anomaly: $v$ and $\hat{a}$ in the present case. For this to be
true, the subset of 
fields must be complete with respect to chiral transformations: Once again,
the Wess-Zumino consistency conditions \cite{WZ} ensure that this is the case.

\end{appendix}

\end{document}